# Organic photovoltaics without p-n junctions: Computational study of ferroelectric columnar molecular clusters

Andrzej L. Sobolewski

## Abstract

Structural and electronic properties of ferroelectric columnar clusters constructed from benzene-1,3,5-tricarboxylic acid, (B3CA)n, were investigated at the Hartree-Fock level. It is shown that B3CA stacks form helix-shaped molecular tubes which are stabilized by intermolecular hydrogen bonds. It is furthermore shown that the strong electric field generated by the uniaxial alignment of the carboxyl groups can split an optically prepared exciton into an electron-hole pair and can drive the charge carriers to the opposite ends of the tube. Some consequences of the phenomenon for photovoltaic applications are discussed.

## Introduction

In the field of photovoltaics, in which manufacturing cost is essential, organics that can be solution-processed at low temperatures offer great advantages compared to their inorganic counterparts. These organic electronic materials generally have significantly lower material costs and can be prepared as thin films using inexpensive room-temperature processes. Organic materials are intrinsically compatible with flexible substrates and high-throughput, roll-to-roll manufacturing processes. Hence, they are ideally suited for the fabrication of large-area electronic and optoelectronic devices with low cost (for a review, see [1] and references therein). Moreover, the electronic and optical properties of organic semiconductors can be tuned to a large extent by appropriate structural modifications, leading to materials with tailored properties for specific applications [2]. On the other hand, organic photovoltaic (OPV) devices still suffer from several drawbacks, such as (i) low exciton diffusion lengths, (ii) non-radiative recombination of charges, and (iii) low carrier mobilities [3, 4].

The latter drawback can be overcome with a promising class of organic materials – discotic liquid crystals (DLC) [5]. The main feature of DLC is that they form columnar phases, where the planar or near-planar monomers form columnar stacks and the columns are packed in a parallel fashion forming a two-dimensional regular lattice. The π-orbital overlap between adjacent aromatic monomers favours a one-dimensional charge transport along the columns. A charge carrier mobility as high as 1 cm2V-1s-1 was observed in the DLC phase of derivatives of hexabenzocoronene [6] and triindole [7, 8]. This value is of the same order of magnitude as the charge carrier mobility of highly ordered single-crystalline rubrene [9] and amorphous silicon. Furthermore, the self-assembly of these materials on the nanometer scale can be manipulated by using, for example, alignment layers, resulting in materials with a monolithic structure on macroscopic length scales that can be produced on a large scale with low cost [10, 11, 12]. Self-organization of liquid crystalline conjugated materials has been used to create thin films directly from solution with structures optimized for use in photovoltaics [13, 14].

Some other weaknesses of OPV devices mentioned above seem to have been overcome in a promising class of new organic and organo-inorganic ferroelectric materials – perovskites are the most celebrated recent examples (for reviews, see [15-21] and references therein). Although the physical mechanism of the photovoltaic effect in organic ferroelectrics is not completely understood, a widely accepted explanation is schematically illustrated in Scheme. 1. The photogenerated electrons and holes are driven in opposite directions by the polarization-induced internal electric field toward the cathode and anode, respectively, and thus contribute to the photovoltaic output. Therefore, the photovoltaic effect in ferroelectrics is essentially different from the conventional interfacial photovoltaic effect in semiconductors, such as p-n junctions. In the junction-based photovoltaic effect, the internal electric field exists only in a very thin depletion layer at the interface. Without the internal field in the bulk material, the photogenerated charge carriers swept out of the depletion region have to diffuse to the cathode or anode. Thus the charge transport is limited by diffusion in junction-based photovoltaics. For the ferroelectric photovoltaic effect, in contrast, a permanent dipole moment and polarization-induced internal electric field exist over the whole bulk region of the ferroelectric, rather than just near a thin interfacial layer. In this case, the charge transport is not limited by diffusion and the output photovoltage is not limited by any energy barrier at the junction.

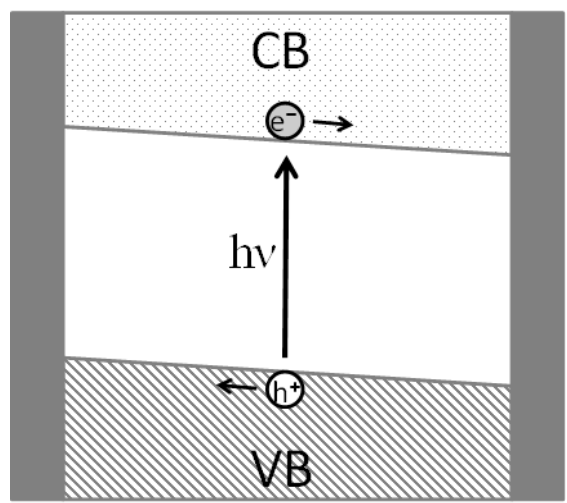

Scheme 1. Model illustrating the mechanism of photovoltaic effect in ferroelectric layer of a semiconductor.

Ferroelectricity and related properties of hydrogen-bonded systems have been studied in many inorganic salts [22]. The proton-transfer and dielectric properties of some hydrogen-bonded molecular crystals have been examined to explore organic analogues of ferroelectric inorganic crystals [23]. Hydrogen bonding also provides an elegant method to enhance the attractive interactions between the aromatic cores of discotic mesogens [24-28]. Moreover, hydrogen bonds which hold the mesogenic units of DLC together may provide the system with an additional feature, namely that each of the subunits has a permanent dipole moment, whose direction is perpendicular to the ring plane. As the molecules stack, the dipoles sum up to yield an exceptionally large dipole moment, thus resulting in the formation of a 'classical' ferroelectric phase. In addition, the assembly of these rods can be influenced by external electric fields due to a field-induced proton-transfer reaction between the molecular subunits [29-30].

In this paper, columnar clusters of carboxy-substituted benzene, the model compound of ferroelectric DLC phases, are studied at the Hartree-Fock (HF) level of electronic structure theory. Additionally, for smaller clusters, some predictions emerging from the HF calculations are checked at an electron-correlated level of theory.

## Results

The molecular subunit of the columnar clusters studied this work is shown in Scheme 2. The central benzene ring represents the aromatic core, while the symmetrically substituted carboxylic groups stabilize the columnar structure due to the formation of intermolecular hydrogen bonds. The polar carboxyl groups are the source of an internal electric field along the molecular stack when the monomers are properly aligned. In real life, the formation of the DLC requires aliphatic side chains (R = CnH2n+1) to prohibit crystallization and the formation of in-plane intermolecular hydrogen bonds. In the present calculations, these side chains were replaced by hydrogen atoms.

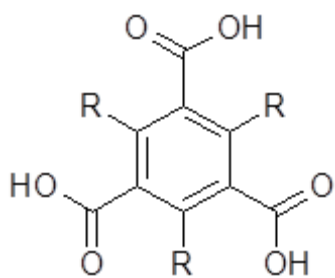

Scheme 2. Chemical structure of compound studied in this work.

The ground-state geometry optimization of the columnar cluster of benzene-1,3,5-tricarboxylic acid (B3CA) results in the formation of a helix-shaped tube having C3-symmetry. The resulting structure and the HF molecular orbitals of the (B3CA)6 cluster are shown in Fig. 1, where the highest 12 occupied (left panel) and the lowest 12 unoccupied (right panel) molecular orbitals are ordered according to their HF energy. Inspection of the figure shows that any clear pattern related to electron (LUMOs) or holes (HOMOs) location can hardly be noticed. One can notice, however, that orbitals are fairly localized on particular molecular units of the cluster with only minor distribution on adjacent units. Thus each orbital can easily be assign to a particular molecular unit. Arranging orbitals vertically according to their HF energy and horizontally according to the location of the electron/hole density on the molecular unit of the cluster results in the pattern plotted in Fig. 2. Interpretation of the pattern is straightforward; HOMO/LUMO localized on a given molecular unit of the cluster represent the top/bottom of the valence/conduction band of the one-dimensional molecular stack. Such orbitals which are connected by a dashed line in Fig. 2 are called 'frontier' orbitals hereafter.

| *Orbital* E[eV] | HOMOs | *Orbital* E[eV] | LUMOs |
|---|---|---|---|
| *108e* -9.603 | 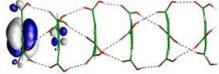 | *112a* 3.199 | 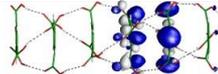 |
| *107e* -10.595 | 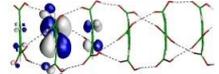 | *116e* 3.151 | 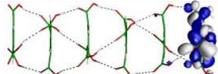 |
| *106e* -11.279 | 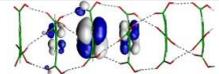 | *111a* 2.721 | 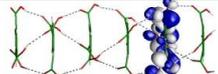 |
| *105e* -11.691 | 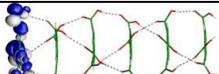 | 115e 2.509 | 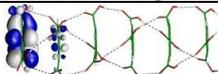 |
| *104e* -11.867 | 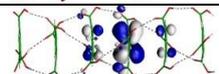 | *114e* 2.430 | 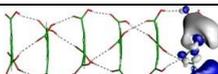 |
| *108a* -12.053 | 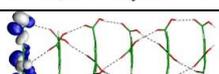 | *110a* 2.056 | 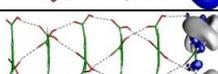 |
| *107a* -12.377 | 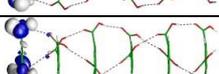 | *109a* 1.960 | 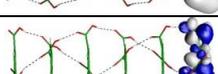 |
| *103e* -12.480 | 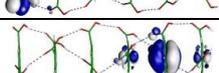 | *113e* 1.612 | 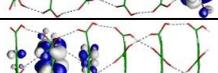 |
| *102e* -12.545 | 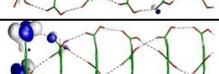 | *112e* 0.945 | 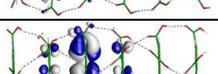 |
| *101e* -12.810 | 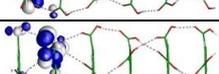 | *111e* 0.357 | 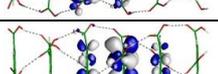 |
| *106a* -13.157 | 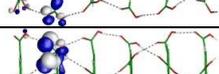 | 110e -0.264 | 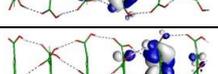 |
| *100e* -13.244 | 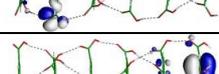 | *109e* -1.057 | 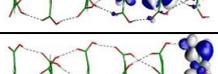 |

Fig. 1. Highest occupied molecular orbitals (HOMOs) and lowest unoccupied molecular orbitals (LUMOs) of a columnar (B3CA)6 cluster computed at the HF/def-SV(P) level. Numbers denote symmetry labels within the C3 point group and orbital energy.

Due to orbital localization, an optical excitation of the stack 'prefers' a local (vertical) excitation of a particular molecular unit. If some higher than LUMO or lower than HOMO orbitals of a given molecular unit are populated with electrons or holes, respectively, they are expected to decay quickly to the respective local LUMO or HOMO. At the top/bottom of the valence/conduction band (VB/CB), electron/hole subjects to a force resulting from interaction with orbitals localized on adjacent molecular units. The force provides unidirectional and oppositely directed energy gradients for electrons and holes which is the essence of the photovoltaic effect in ferroelectrics (c.f. Scheme 1). Interestingly, the frontier HOMOs and LUMOs visualized in Fig. 2 are doubly degenerate, having e symmetry in the C3 point group, and are thus subject to the Jahn-Teller effect when singly occupied with an electron or a hole.

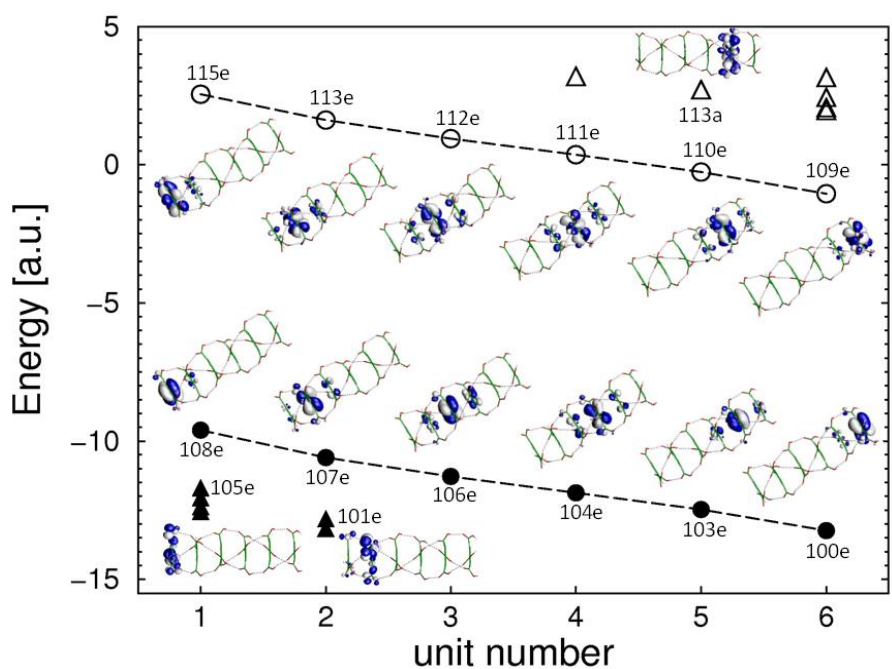

Fig. 2. HOMOs (full symbols) and LUMOs (empty symbols) of a columnar (B3CA)6 cluster plotted according to their HF energy and location of the center of electron/hole density assigned to a particular molecular units. Numbers denote orbital symmetry within the C3 point group. Frontier orbitals are denoted by circles and connected by dashed line.

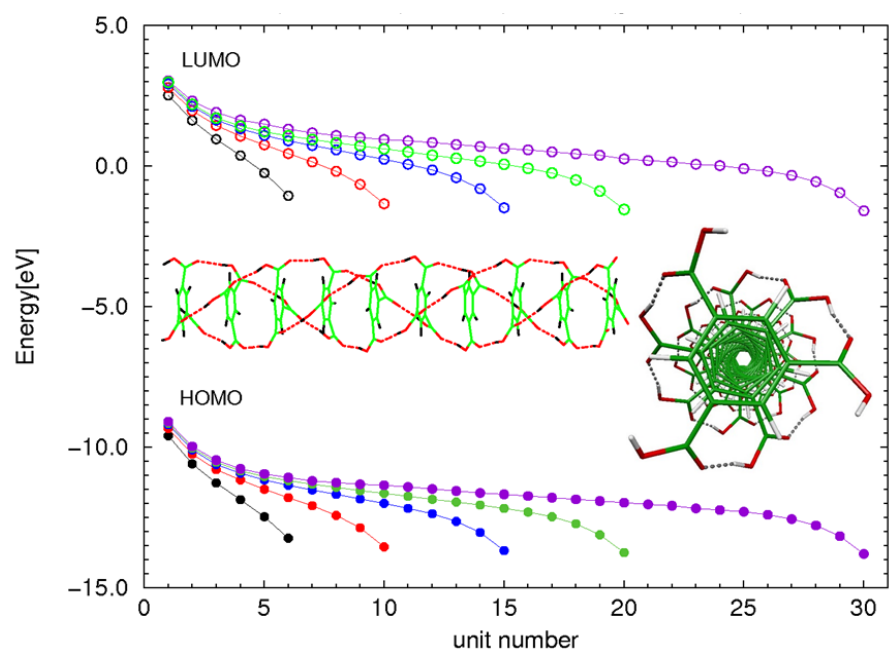

Fig. 3. Energy of frontier HOMOs and LUMOs of the columnar (B3CA)N clusters computed at the HF/def-SV(P) level and plotted vs. their localization on successive molecular units for N=6 – black, N=10 – red, N=15 – blue, N=20 – green, and N=30 – violet.

Elongation of the molecular stack (N = 6-30, see Fig. 3) results in a stretching of the molecular energy gradients along the column, while the electric driving force given by the energy difference between the terminal occupied orbital (holes) and the terminal unoccupied orbital (electrons) saturates at a constant value with increasing cluster size (see Fig. S1 of ESI).

| E[eV] | HOMOs | E[eV] | LUMOs |
|---|---|---|---|
| *180e* -9.335 | 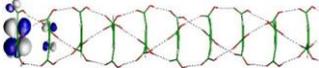 | *191e* 2.782 | 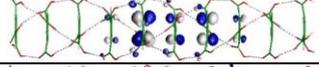 |
| *179e* -10.253 | 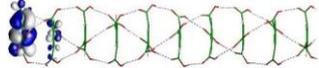 | *189e* 1.961 | 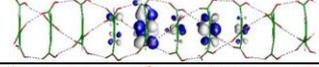 |
| *178e* -10.804 | 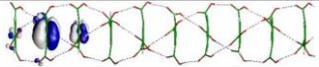 | *188e* 1.433 | 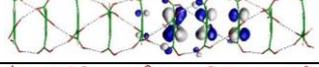 |
| *177e* -11.191 | 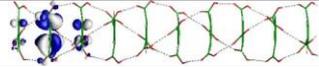 | *187e* 1.052 | 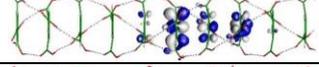 |
| *175e* -11.506 | 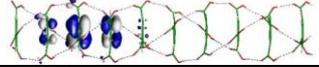 | *186e* 0.736 | 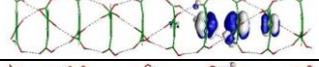 |
| *174e* -11.796 | 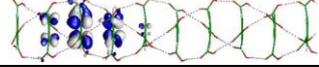 | *185e* 0.441 | 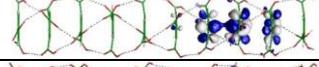 |
| *173e* -12.092 | 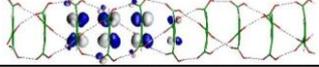 | *184e* 0.140 | 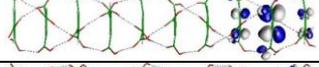 |
| 171e -12.430 | 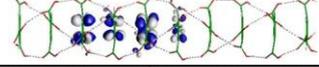 | *183e* -0.203 | 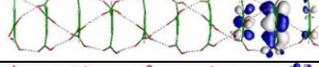 |
| *169e* -12.878 | 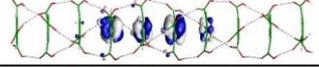 | *182e* -0.659 | 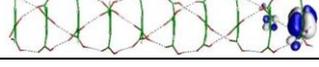 |
| *165e* -13.547 | 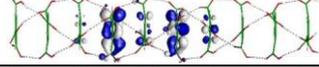 | *181e* -1.354 | 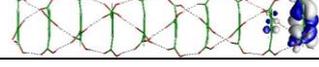 |

Fig. 4. Frontier Hartree-Fock HOMOs and LUMOs of the columnar (B3CA)10 cluster computed at the HF/def-SV(P) equilibrium geometry of the ground state.

Elongation of the molecular stack (N = 10, Fig. 4) results in an increased orbital delocalization, particularly in the central region of the cluster. The effect is even more pronounced for longer tubes (N = 15 - Fig. S2 and N = 20 - Fig. S3 of ESI). This effect can be related to the homogeneity of the electric field inside the tube which leads to constant orbital energy gradient (Fig. 3). Orbital (de)localization along the tube results from the ordered structure of the stacks and indicates a significant mobility of charge carriers along the tube.

To answer the question how orbital (de)localization along the stack is dependent on the computational method, in Fig. S4 of the ESI the frontier HOMOs and LUMOs determined with the aid of the DFT/B3LYP/def-SV(P) method at the HF/def-SV(P) equilibrium geometry in the ground state of the (B3CA)10 cluster are shown. One can notice upon inspection of the figure that the DFT orbitals are qualitatively similar to the respective HF orbitals of Fig. 4. Interestingly, apart from the energy difference which is method-dependent, a general shape of the top of the valence band (HOMOs) and the bottom of the conduction band (LUMOs) is apparently similar for both theoretical methods (c.f. Fig. S5 of ESI).

An interesting question is whether the internal electric field in such molecular stacks is large enough to split an exciton into a pair of charge carriers (electron and hole) and to drive them to the opposite ends of the tube thus generating a photovoltaic effect. To answer this question, equilibrium nuclear geometry of the (B3CA)10 cluster in the lowest excited state (triplet) was optimized at the unrestricted Hartree-Fock (UHF) level. The resulting UHF natural orbitals which are singly occupied in the triplet state, are shown in Fig. 5. They are localized at the opposite ends of the tube, thus clearly indicating the spontaneous splitting of the exciton into charge carriers and the separation of the charge carriers under the action of the internal electric field.

SOMO1            SOMO2

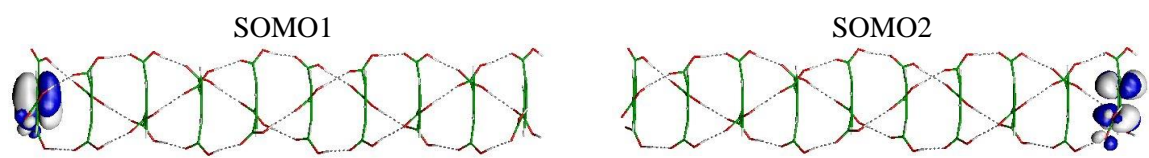

Fig. 5. Singly-occupied natural molecular orbitals of the lowest excited triplet state of the columnar (B3CA)10 cluster computed at the UHF/def-SV(P) level.

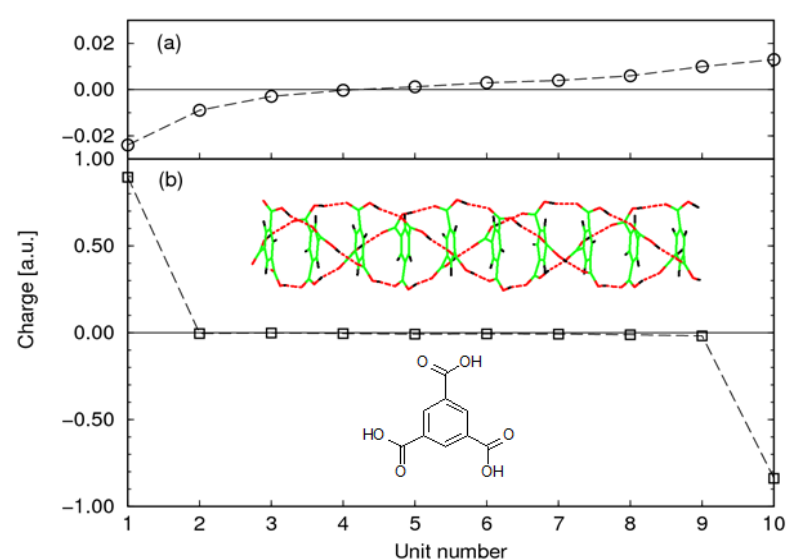

Fig. 6. Distribution of charge on particular molecular units of the columnar (B3CA)10 cluster in the ground state (a) and in the lowest excited triplet state (b).

To support further this finding, the population analyses for the ground state and for the lowest excited triplet state, based on occupation numbers yielding "shared electron numbers" and multicenter contributions implemented in Turbomole [36] are shown in Fig. 6. The atomic charges were summed up for each monomer. One can see that the molecular stack is already strongly polarized in the ground state with a huge dipole moment (112.8 Debye) aligned along the C3 symmetry axis. Electronic excitation shifts electronic charge from the electronegative end of the stack to the electropositive end and reverses the sign of the dipole moment (-29.3 Debye). The charge that localizes on the terminal molecular units of the stack upon electronic excitation is close to unity with only minor population of other units.

While the lowest triplet state was analyzed as an elementary electronic excitation for technical reasons, electronically 'relaxed' singlet and triplet excited states are essentially degenerate in these systems. The clear spatial separation of the singly occupied orbitals (see Fig. 5) indicates a biradicalic electronic structure, where triplet and singlet states are degenerate. The charge-separated state can thus spontaneously be formed by optical excitation within the singlet manifold of the (B3CA)10 cluster. As mentioned above, the frontier HOMOs and LUMOs of the columnar (B3CA)n cluster are doubly degenerate in the C3 symmetry point group. Thus these systems are subject to the Exe Jahn-Teller effect (JTE) upon electronic excitation. This effect may stabilize an exciton and, in the case of a dynamical JTE, may increase charge carrier mobility along the stack due to additional coupling between vector potentials of the adjacent molecular units.

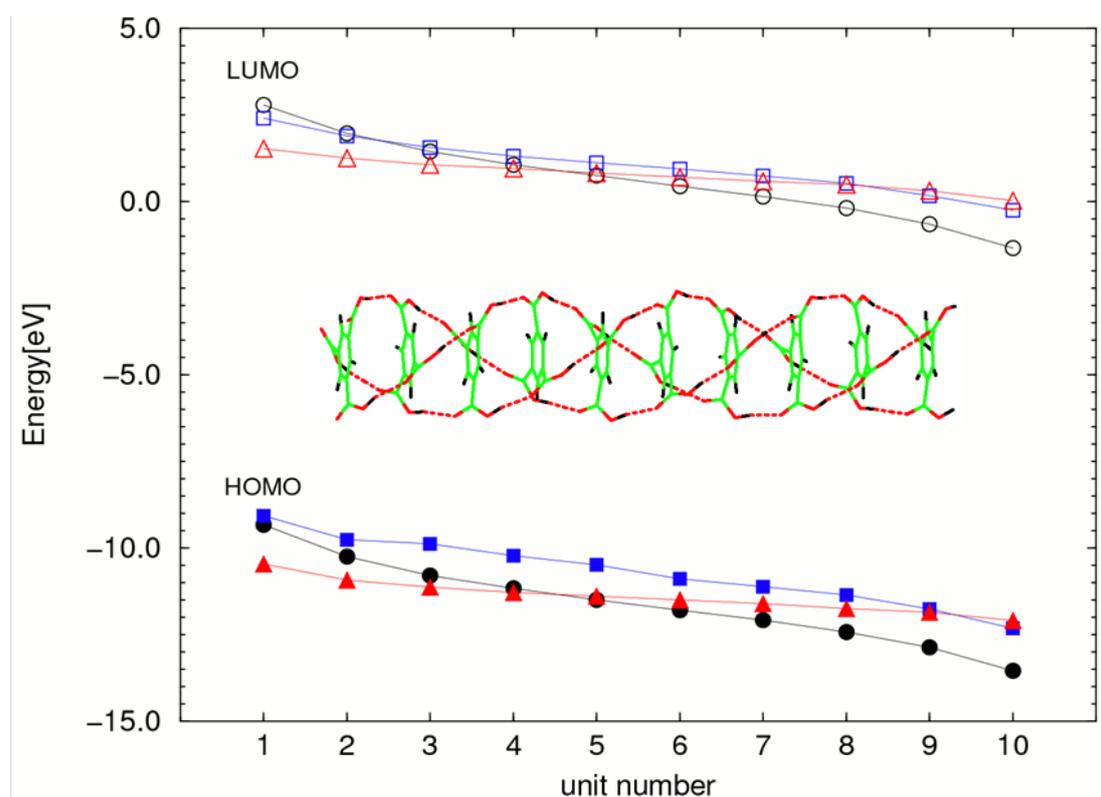

To further explore the question of the relationship of the excited-state charge separation to the Fig. 7. Frontier HOMOs and LUMOs of the columnar (B2CA)10 cluster – blue squares, partially polarized (B3CA)10 cluster – red triangles, and the fully polarized (B3CA)10 cluster – black circles, computed at the HF/def-SV(P) level.

strength of the internal electric field, two clusters which are less polar than the (B3CA)10 stack were studied; (i) the (B2CA)10 cluster, composed of benzene-1,4-dicarboxylic acid units, and (ii) the (B3CA)10 cluster, where one of the hydrogen-bonded carboxylic chains is oriented in the opposite direction with respect to the two other chains. These systems have significantly reduced dipole moments of the ground state (77.1 and 39.5 Debye, respectively) compared to 112.8 Debye of the uniaxially aligned (B3CA)10 cluster. The resulting distribution of the frontier orbital energies for these systems is shown in Fig. 7. One can see that the internal orbital-energy gradient, which governs the photoinduced charge separation, is directly related to the polarity of the system.

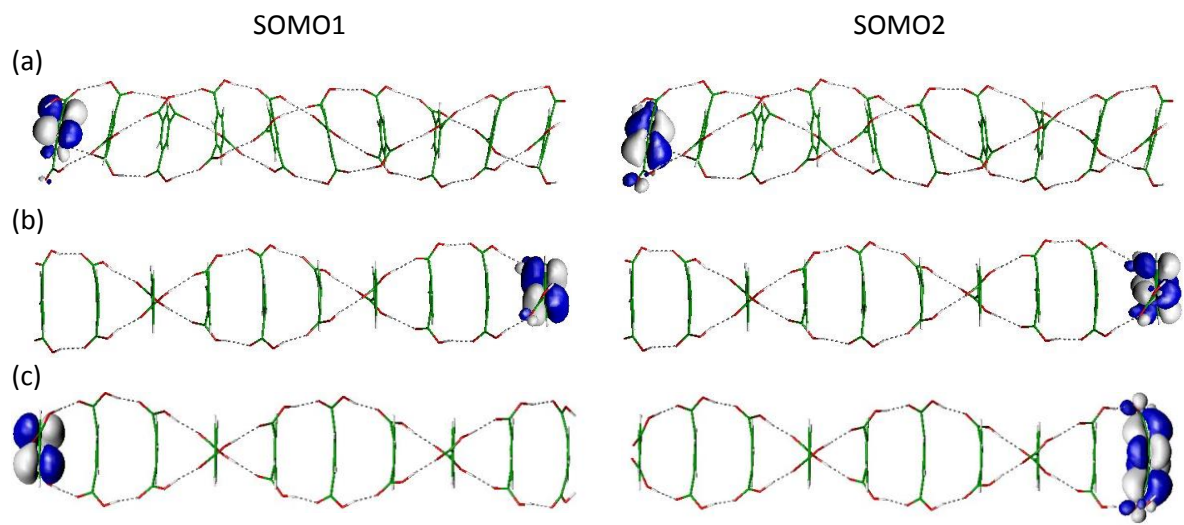

Fig. 8. Singly-occupied natural molecular orbitals in the T1 state of partially polarized (B3CA)10 cluster (a), and in the T1 state (b) and in the S1 state (c) of the (B2CA)10 cluster computed at the UHF/def-SV(P) level.

Geometry optimization of the lowest excited triplet state in these less polar systems shows that their internal electric field is too weak to split an exciton into an electron-hole pair in the lowest excited (triplet) state. This is clearly seen from inspection of the singly-occupied UHF natural orbitals (Figs. 8a and 8b) as well as from the analysis of the charge distribution over the molecular units of the stacks (Figs. 9a and 9b). Fig. 8 reveals, moreover, that the T1 state excitons are localized on the terminal molecular units of the clusters, but on opposite ends: at the electro-positive end of the (B2CA)10 cluster and at the electro-negative end of the partially aligned (B3CA)10 cluster. These examples clearly show that there exists a threshold value the of internal electric field which is related to the dipole moment of the monomer, below which the field is too weak to overcome the Coulomb attraction between the charge carriers. Interestingly, geometry optimization of the lowest excited singlet state of the (B2CA)10 cluster resulted in splitting of the exciton and in separation of charges on the opposite ends of the cluster (Figs. 8c and 9c). Since for the biradical structure (Fig. 8c) singlet and triplet states are essentially degenerate, this result clearly shows that for small orbital-energy gradient there is a competition between exciton splitting in the energetically higher singlet manifold and exciton localization in the energetically lower triplet manifold, the former resulting in generation of a photovoltaic effect and the latter in trapping of exciton leading eventually to geminate recombination.

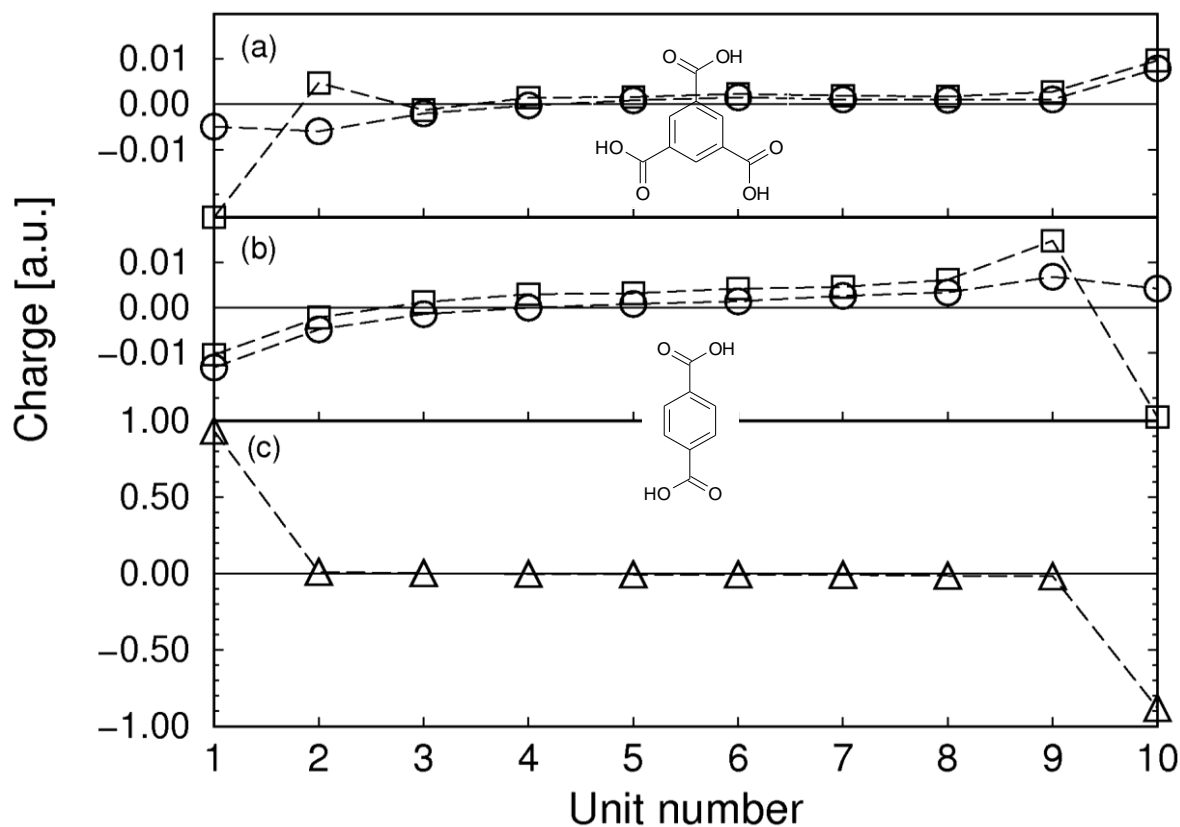

Fig. 9. Distribution of charge on particular molecular units of the columnar partially polarized (B3CA)10 -(a) and (B2CA)10 - (b) clusters in the ground state (circles), in the lowest excited triplet state (squares), and in the lowest excited singlet state of the (B2CA)10 cluster – (c).

In order to verify the predictions of the model obtained at the HF and DFT levels of theory, the algebraic diagrammatic construction method of second order, ADC(2), was used for the computation of the lowest electronically excited states. With present hardware limitations, such computations are only possible for smaller clusters of this molecular system. Additionally, the current implementation of the method cannot handle non-abelian symmetry point groups. Therefore, the C3 symmetry of the (B3CA)n cluster cannot be exploited at this theoretical level. The largest system for which it was possible to obtain converged results at the ADC(2) level was the (B2CA)3 cluster treated in the C2 point group.

The lowest singlet and triplet vertical excitation energies of the (B2CA)3 cluster computed with this method are collected in Table S1 of the ESI. The relevant frontier Hartree-Fock molecular orbitals are displayed in Table S2 of the ESI. Inspection of these results shows that the lowest excited singlet and triplet states are of $\pi\pi^*$ electronic nature and involve a partial transfer of electron density from the electro-negative end of the cluster to the electro-positive one. This is manifested as an decrease of the dipole moment compared to the ground state. Optical excitation of the lowest singlet states is only a weakly allowed, which is typical for aromatic systems of this size.

An interesting result is obtained in the course of the ADC(2) geometry optimization of the lowest triplet and singlet excited states of the system. First of all, a remarkable energetic stabilization (by about 0.5 eV) of the states is obtained (c.f. Table S1 of ESI and Table 1). Secondly, geometry optimization of the S1 state results in almost complete splitting of the exciton and localization of charges on the opposite ends of the cluster (Fig. 10). This is not the case, however, in the lowest excited triplet state. The wavefunction of this state conserves the locally-excited character with only a small shift of the charge (Fig. 11) similar to that determined at the geometry of the ground state (Tables S1 and S2 of ESI). The photophysical scheme of the B2CA stack which emerges from the ADC(2) study of the (B2CA)3 cluster is presented in Fig. 12. This result is in full accord with results obtained at the UHF level for the (B2CA)10 cluster (Figs. 8b and 8c) and justifies the qualitative discussion of the phenomena for larger clusters at this relatively simple theoretical level.

Table 1. Adiabatic excitation energies of the lowest excited singlet and triplet states (ΔE), energies of 'vertical' emission from these states, oscillator strength (f), dipole moment (µ), and the most important electronic configurations computed at the ADC(2)/cc-pVDZ-optimized geometry of the (B2CA)3 cluster in a given electronic state.

| State | ΔE/eV | f | µ/Debye | Electronic Configuration |
|---|---|---|---|---|
| $^1$CT | 3.45[a] | - | 3.64 | 0.82(66b-64a) |
| S$_0$($^1$CT) | 2.20[b] | 0.0032 | 18.53 | (63a)$^2$(66b)$^2$ |
| $^3$LE | 3.05[a] | - | 12.33 | 0.79(66b-64a) −0.36(64b-64a) +0.32(66b-65a) |
| S$_0$($^1$LE) | 2.34[c] | - | 14.07 | (63a)$^2$(66b)$^2$ |

[a] adiabatic energy, relative to the minimum of the ground state.
[b] energy of the 'vertical' fluorescence.
[c] energy of the 'vertical' phosphorescence.

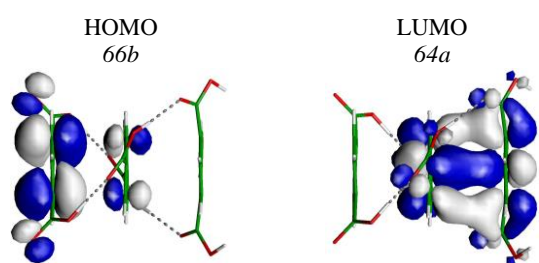

Fig. 10. Hartree-Fock HOMO and LUMO computed at the equilibrium geometry of the S1 state of the (B2CA)3 cluster with symmetry labels indicated.

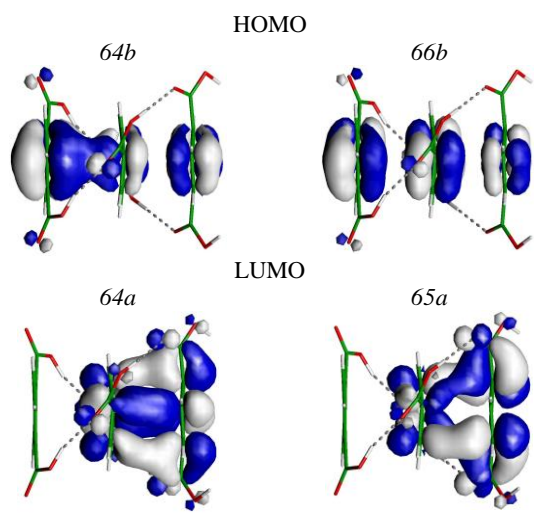

Fig. 11. Hartree-Fock HOMOs and LUMOs computed at the equilibrium geometry of the T1 state of the (B2CA)3 cluster with symmetry labels indicated.

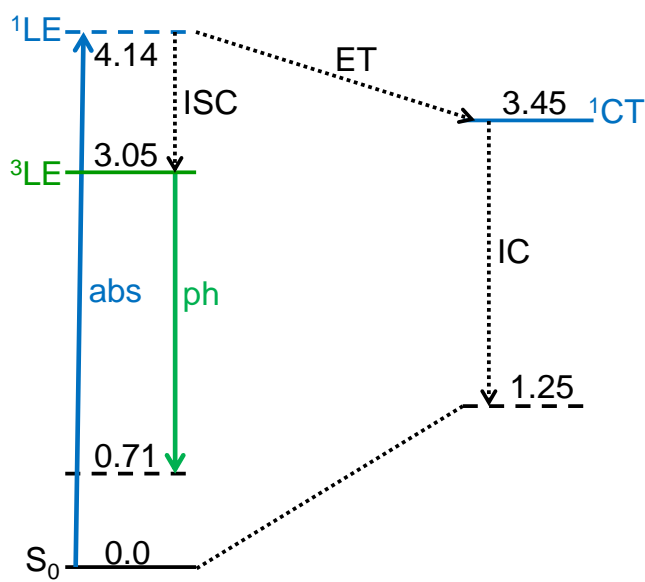

Fig. 12. Photophysical scheme of the (B2CA)3 cluster determined at the ADC(2)/cc-pVDZ level of theory. Solid/ dashed lines denote adiabatic/vertical energy of a given electronic state. Solid/dotted arrows denote radiative/ radiationless processes. Numbers denote energy in eV.

## Discussion

Ab initio explorations of the structure of model ferroelectric molecular systems have shown that they form fairly rigid helix-shaped molecular nano-tubes due to the stacking interaction of aromatic cores and intermolecular hydrogen bonds between the molecular units. The polar monomers, if properly aligned, generate an internal electric field which can potentially break an optically generated exciton into an electron-hole pair and drive these charge carriers to the opposite ends of the tube. A lattice composed from such molecular stacks may form a ferroelectric DLC phase that possesses promising properties with respect to photovoltaic applications.

There are several conditions to be fulfilled for potential applications of such ferroelectric molecular stacks in photovoltaics:

(i) The systems have to absorb in the visible range of the spectrum; This condition can be fulfilled by an extension of the aromatic π-system of the monomer unit and/or chemical substitutions of the aromatic core (work in progress).

(ii) There exists a limiting value of the internal electric field strength below which optically generated excitons do not split spontaneously into charge carriers; In the model systems studied in this work, consisting of benzene cores augmented with carboxyl groups as molecular dipoles, two uniaxially aligned groups are necessary to achieve the effect in the lowest excited singlet state.

(iii) he system has to be photostable; Generally, fused aromatic rings like coronene, hexabenzocoronene, etc., which are commonly used as monomers of the DCLs, are intrinsically highly photostable. Another problem is the photostability of the ferroelectric phase of the system. Since the ferroelectricity of the systems studied in the present work is related to the polarity of hydrogen-bonded carboxyl groups, any thermally- or photo- induced proton-transfer reaction along a hydrogen-bonded chain can compromise this property. This issue is discussed in more detail below.

The stability of the ferroelectric phase of a hydrogen-bonded molecular stack depends on which system is more stable: the one with the highest polarity (all hydrogen bonds oriented in the same direction) or that with lower polarity, in which some hydrogen-bond chains are oriented in opposite direction? Geometry optimization of a model (B3CA)10 stack performed at the HF level shows that the less polar system, in which one of three hydrogen-bond chains is oriented in the opposite direction, is energetically preferred. One has to keep in mind, however, that these calculations were performed at relatively simple (uncorrelated) level and refer to the system in a vacuum. Inter-column interactions in the DLC lattice composed of such ferroelectric stacks may be decisive in this respect. Another observation made in calculations is the increase of the polarity per molecular unit with increasing cluster size. Although the average dipole moment of a single B3CA unit seems to saturate with increasing of the cluster size (see Fig. S6), it still increases by about 10% from $N = 10$ to $N = 30$. Thus for systems of a real nano-size the field-induced collective polarity may constitute an important factor.

Another possibility to avoid depolarization by proton transfer reactions is the construction of ferroelectric DLC structures which are not based on hydrogen bonding. An example of a polar moiety which does not form hydrogen bonds is an aromatic core (benzene) augmented with acetonitrile side groups - 1,3,5-benzene-triacetonitrile (B3CN). The frontier orbital energies of the columnar (B3CN)10 cluster are shown in Fig. 10 together with the data for the unipolar (B3CA)10 cluster. It can be seen that the replacement of the carboxy groups by acetonitrile groups does not lead to qualitative changes of the effect. The average dipole moment per molecular unit is almost the same in both clusters (11.3 Debye in (B3CA)10 vs. 11.8 Debye in (B3CN)10). However, this structural modification significantly changes the stacking distance of benzene rings which increases from 3.92 Å for the carboxy-substituted monomer to 5.24 Å for the acetonitrile-substituted monomer. The increased distance of the aromatic cores decreases the intermolecular coupling and results in the localization of the molecular orbitals on single monomers of the cluster (see Fig. S7 of ESI). On the other hand, the driving force for charge separation does not change substantially (Fig. 10) and is strong enough to split an exciton and localize the resulting charge carriers on the opposite ends of the stack (see Figs. S8 and S9 of ESI).

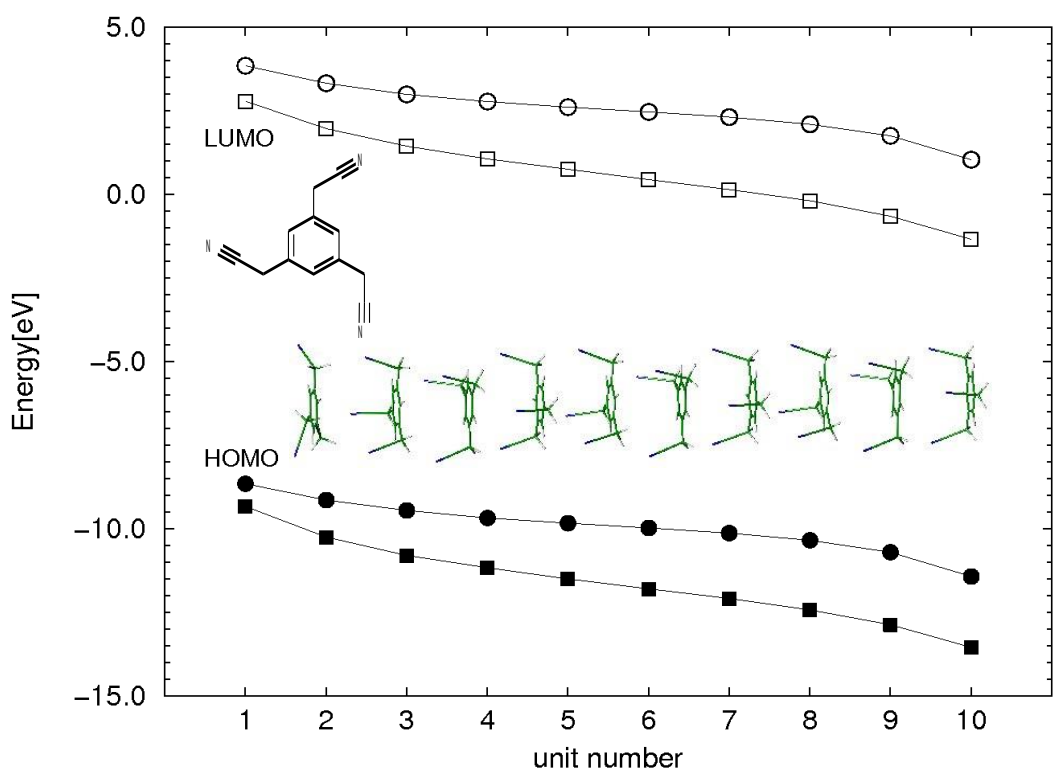

Fig. 10. Energy of the frontier HOMOs and LUMOs of the columnar (B3CN)10 cluster – circles, and (B3CA)10 cluster – squares, computed at the HF/def-SV(P) level.

Conclusions

The theoretical explorations of the structures and electronic properties of novel ferroelectric molecular systems performed in this study result in the following conclusions:

(i) Stacked hydrogen-bonded BnCA aromatic cores form rigid helix-shaped columnar assemblies.

(ii) Strong hydrogen bonds in (BnCA)N column give rise to short stacking distances, resulting in strong π-stacking interactions and potentially high charge carrier mobility along the column.

(iii) The exceptionally high polarity of the columnar assemblies provides a driving force for the splitting of excitons and the separation of the resulting charge carriers (electrons and holes), resulting in their localization on opposite ends of the assemblies and generating a photovoltaic effect.

These properties of ferroelectric columnar assembles make them very attractive candidates for organic photovoltaic applications.

## Theoretical methodology

The equilibrium geometries in the electronic ground state and in the lowest excited triplet and singlet states of larger clusters were optimized at the Hartree-Fock level of theory. For ground-state geometry optimization of smaller clusters the second-order Møller-Plesset (MP2) method was used. The ADC(2) method (algebraic diagrammatic construction of second order) [31] was used for optimization of the excited-state geometries and for the computation of electronic properties. This methodology offers a description of electronically excited states which is of similar quality as is the MP2 level for the electronic ground state. The RI (resolution of identity) approximation was employed in the MP2 and ADC(2) calculations [32]. The correlation-consistent split-valence double-zeta basis set with polarization functions on all atoms (cc-pVDZ) [33] was used in these calculations, while in the calculations performed at the Hartree-Fock (HF) and Density Functional Theory (DFT) levels a more compact basis set (def-SV(P)) [34] was utilized. In DFT calculations the Becke, three-parameter, Lee-Yang-Parr (B3LYP) functional [35] was used. All calculations were carried out with the Turbomole program package [36].

## Acknowledgements

This work was supported by the National Science Center of Poland (Grant No. 2012/04/A/ST2/00100). I thank Wolfgang Domcke for helpful comments on the manuscript.